\documentclass[prb, amsmath,amssymb,preprint, showpacs,showkeys,superscriptaddress]{revtex4-1}

\usepackage{graphicx,epstopdf,subfigure,color}
\usepackage{dcolumn}
\usepackage{bm}
\usepackage{amsmath,amssymb}
\usepackage{ulem}  

\usepackage[]{hyperref}
\hypersetup{
    colorlinks,%
    citecolor=blue,%
    filecolor=black,%
    linkcolor=blue,%
    urlcolor=black
}

\begin{document}
\newcommand{\dV}[0]{\, \text{d}V}
\newcommand{\Gc}[0]{\, G^\text{c}}
\newcommand{\Fc}[0]{\, \hat{G}^\text{c}}
\newcommand{\Gg}[0]{\, G^\text{g}}
\newcommand{\Fg}[0]{\, \hat{G}^\text{g}}

\title{Atomistically-enabled non-singular anisotropic elastic representation of near-core dislocation stress fields in $\alpha$-iron}

\author{Dariush Seif}
\address{Fraunhofer-Institut f\"{u}r Werkstoffmechanik IWM, W\"ohlerstra{\ss}e 11, 79108 Freiburg, Germany}
\author{Giacomo Po}
\address{University of California Los Angeles, 405 Hilgard Avenue, Los Angeles, CA 90095, USA}
\author{Matous Mrovec}
\address{Fraunhofer-Institut f\"{u}r Werkstoffmechanik IWM, W\"ohlerstra{\ss}e 11, 79108 Freiburg, Germany}
\author{Markus Lazar}
\address{Heisenberg Research Group, Department of Physics, Darmstadt University of Technology, Hochschulstra{\ss}e 6, D-64289 Darmstadt, Germany}
\author{Christian Els\"asser}
\address{Fraunhofer-Institut f\"{u}r Werkstoffmechanik IWM, W\"ohlerstra{\ss}e 11, 79108 Freiburg, Germany}
\author{Peter Gumbsch}
\address{Fraunhofer-Institut f\"{u}r Werkstoffmechanik IWM, W\"ohlerstra{\ss}e 11, 79108 Freiburg, Germany}
\address{Institute for Applied Materials (IAM), Karlsruhe Institute of Technology, Kaiserstra{\ss}e 12, 76131 Karlsruhe, Germany}

\date{\today}

\begin{abstract}
The stress fields of dislocations predicted by classical elasticity are known to be unrealistically large approaching the dislocation core, due to the singular nature of the theory. While in many cases this is remedied with the approximation of an effective core radius, inside which ad hoc regularizations are implemented, such approximations lead to a compromise in the accuracy of the calculations. 
In this work, an anisotropic non-singular elastic representation of dislocation fields is developed to accurately represent the near-core stresses of dislocations in $\alpha$-iron. The regularized stress field is enabled through the use of a non-singular Green's tensor function of Helmholtz-type gradient anisotropic elasticity, which requires only a single characteristic length parameter in addition to the material's elastic constants. Using a novel magnetic bond-order potential to model atomic interactions in iron, molecular statics calculations are performed, and an optimization procedure is developed to extract the required length parameter. Results show the method can accurately replicate the magnitude and decay of the near-core dislocation stresses even for atoms belonging to the core itself. Comparisons with the singular isotropic and anisotropic theories show the non-singular anisotropic theory leads to a substantially more accurate representation of the stresses of both screw and edge dislocations near the core, in some cases showing improvements in accuracy of up to an order of magnitude. The spatial extent of the region in which the singular and non-singular stress differ substantially is also discussed. The general procedure we describe may in principle be applied to accurately model the near-core dislocation stresses of any arbitrarily shaped dislocation in anisotropic cubic media. 
\end{abstract}

\maketitle

\section{Introduction\label{sec_Introduction}}

The response of dislocations to complex short and long-range internal stresses has long been known to govern a wide array of macroscopic material phenomena, including basic mechanical properties. Such stresses may arise from external loading, or by interactions with internal material defects, such as neighboring dislocations, point-defects, and other material interfaces. 
Investigations of the collective behavior of discrete dislocation ensembles by dislocation dynamics (DD) methods \cite{Lepinoux_1987, Ghoniem_1988, Schwarz_1997, Zbib_1998, Weygand_2002, Bulatov_2004} have proven successful in modeling material deformation at the micro-scale. Due to the convenience of having closed-form expressions for the elastic fields, such simulations have primarily focused on isotropic materials. For anisotropic materials, the literature is sparse with such studies, due to the heavy computational requirements of numerical integrations in the elastic fields \cite{Bacon_1978}. While efforts have been made to increase computational efficiency \cite{Yin_2010,Yin_2012}, computational times can still be more than two orders of magnitude longer for anisotropic versus isotropic simulations. Thus, such simulations are currently relegated to very short times and very few dislocations. This has limited large-scale simulations of dislocation ensembles to materials with low anisotropy.

The most common measure of a cubic material's anisotropy is through the Zener anisotropy ratio \cite{Zener_1948}, $A$, which is defined by a material's elastic constants as $A=c_{44}/c^\prime$, where $c^\prime=(c_{11}-c_{12})/2$. The physical interpretation is that of a ratio of the resistance to shear on $\{$100$\}$ planes in $\langle$100$\rangle$ directions ($c_{44}$) to the resistance to shear in $\langle$110$\rangle$ directions ($c^\prime$). Materials having a Zener ratio of unity are taken as isotropic. Iron is a material well known to exhibit significant anisotropy ($A$=2.43) which drastically increases as the $\alpha-\gamma$ phase-transition temperature (912 $^\circ$C) is approached \cite{Dever_1972}. Due to its abundance and favorable mechanical properties, it still remains as the basis of today's most technologically vital materials. Several recent studies have employed classical anisotropic elasticity theory and specifically illustrated the importance of anisotropy in discrete dislocation calculations. Using anisotropic elasticity theory, Dudarev et al. \cite{Dudarev_2008} showed that the relative values of their elastic free energy may account for the anomalous abundance of $a\langle$100$\rangle$ rather than $a/2\langle$111$\rangle$ dislocation loops observed experimentally at high temperatures. Their analysis showed such a conclusion would not be reached via isotropic treatments. Through DD simulations, Aubrey et al. \cite{Aubry_2011} showed the effect of anisotropy on the equilibrium shapes of dislocation shear loops over a wide temperature range. Experimentally observed slip-system-dependent loop features such as sharp corners were able to be reproduced, where not possible using the isotropic theory. Similar DD simulations of Frank-Read sources showed that the isotropic elastic approximation leads to large errors in their activation stresses, further increasing with temperature \cite{Fitzgerald_2012}. Dependence on orientation was shown to play a very important role, specifically in allowing certain configurations much lower critical stresses to become sources than isotropic theories allow, and more aligned with experimental observations. Thus, the modeling of dislocations in $\alpha$-Fe cannot be accomplished via isotropic elasticity. 

While modeling via anisotropic theories has made recent progress, current implementations, including those cited thus far, must still disregard material contributions near the core due to the singular nature of such theories. At the same time, it is well known that many important physical processes undergone by dislocations, including, among others, their intersection, pinning/de-pinning, and capture/emission of point-defects, can be accurately described only in terms of the processes occurring at the sub-nanometer scale adjacent to the dislocation core. To avoid the singularity issue, a core radius \cite{Gavazza_1976} must be prescribed around the dislocation, typically several Burgers' vector magnitudes in radius or larger, within which processes may occur in a pre-defined, ad hoc fashion, losing the atomic-scale resolution. In cases where heightened resolution on the atomic aspects near the core are required, such assumptions are not-acceptable, as a complete model of the complex interactions of dislocations is possible only when all material contributions governing such phenomena are included. 
One example where the near-core field is crucial is in the study of creep in metals. Creep is a phenomenon governed largely by the climb of dislocations in response to their absorption and emission of vacancies \cite{Weertman_1955}. This phenomena is further enhanced as the otherwise isotropic diffusion of vacancies in metals is made anisotropic in the stress fields of dislocations due to drift forces caused by gradients in their mutual interaction energy. This interaction energy may be expressed as the product of the dislocation's stress tensor with the vacancy's (constant) formation volume tensor, as $-\sigma^\perp_{ij}(\textbf{r})V^f_{ij}$. Thus, to capture the absorption process accurately, the stress field near the core must be known at the atomic scale to a high degree of accuracy. If such resolution could be obtained, a lattice kinetic Monte Carlo simulation could be employed to accurately calculate point defect capture rates of arbitrarily shaped dislocations. Such accuracy is not possible via the singular theories, as the predicted stresses become too large and capture radii are required. 

We seek here to develop and employ a computationally-feasible continuum dislocation theory that can account for full material anisotropy, and can provide accuracy down to the atomic scale. While the end goal of a reliable multiscale material model may be obtained by the coupling of continuum to atomistic methods \cite{Srivastava_2013}, such approaches require a large computational overhead and cannot offer the degree of flexibility of a purely continuum-based approach. Recent progress in employing gradient elasticity \cite{Mindlin:1964tg} of the Helmholtz type for the modeling of dislocation fields \cite{Lazar_2012,Lazar_2013,Lazar:2014dy} has shown much promise for the heightened resolution of DD simulations in isotropic materials, as illustrated by Po et al. \cite{Po_2014}. The attractiveness of this approach is that it leads to a non-singular description of the stress and displacement fields of dislocations and requires only one additional material parameter. With the recent derivation of the anisotropic non-singular Green's tensor function of gradient elasticity of Helmholtz type by Lazar and Po \cite{PoLazar_ANSGF_2014}, in this work we explore its application to describing the near-core stress fields of dislocations in anisotropic materials, where singular theories are not valid. Choosing $\alpha$-iron as a model anisotropic material, we perform molecular statics calculations of screw and edge dislocations in $\alpha$-iron using a recently developed magnetic bond-order potential (BOP) by Mrovec et al. \cite{Mrovec_2011}. An optimization scheme is then developed to extract the required length-scale parameter based on the obtained atomic stresses. While the focus here is on the development and verification of a non-singular anisotropic dislocation theory, knowledge of the acquired length parameter allows the non-singular Green's function to be applied to the modeling of other elastic fields, such as those due to point defects. We also note in a more recent work by Lazar and Po \cite{LazarPo_PLA2015}, a method to treat materials with varying degrees of anisotropy beyond cubic was developed, where the Green's function used in the current work is a special case of that more general theory.

In what follows, first a review of the classical anisotropic continuum expressions for dislocation displacements and stresses is given in section \ref{sec_Theory}. We then show that these classical expressions can be made non-singular by the direct substitution of the non-singular anisotropic Green's tensor function. The equations used to calculate the atomic stresses from the BOP relaxations are also described. In section \ref{sec_Calculations}, a description of the atomistic simulations that were performed is given, including the methodology in which the non-singular characteristic length parameter, $\ell$, was extracted from such calculations. Results and Discussion of the atomistic and continuum calculations and of the spatial extent to which the non-singular theory is important are given in sections \ref{sec_Results}, followed by a summary and conclusions from the work in section \ref{sec_Conclusions}.

\section{Stress Fields of Dislocations\label{sec_Theory}}
In this section we derive analytical expressions for the Cauchy stress generated by a dislocation loop in both classical and gradient anisotropic elasticity. These expressions are implemented numerically in section \ref{sec_Results} and compared to atomistic calculations. 

The classical elastic theory of discrete dislocations can be introduced in two equivalent frameworks. The first is framework of compatible elasticity, which was established by Volterra \cite{Volterra_1907} three decades prior to the proposed existence of crystal dislocations \cite{Dehlinger_1929, Taylor_1934,Orowan_1934,Polanyi_1934} and five decades before the  proof of their existence by early transmission electron microscopy \cite{Menter_1956,Hirsch_1956}. In his approach, Volterra studied a compatible elastic problem in which dislocations are represented by  constant translational discontinuities of the displacement field across an internal material surface. In this paper, however, we adopt the alternative  framework of incompatible elasticity originally proposed by Kr\"oner \cite{Kroner:1958aa} and Mura \cite{Mura_1963}. The main kinematic assumption of this framework is that the displacement gradient $u_{i,j}$ is split additively into an elastic distortion $\beta^E_{ij}$ and a plastic distortion $\beta^P_{ij}$:
\begin{align}
u_{i,j}=\beta^E_{ij}+\beta^P_{ij}\, .
\label{linear_decomposition}
\end{align}
\begin{figure}[t]
   \centering
   \includegraphics[width=.7\textwidth]{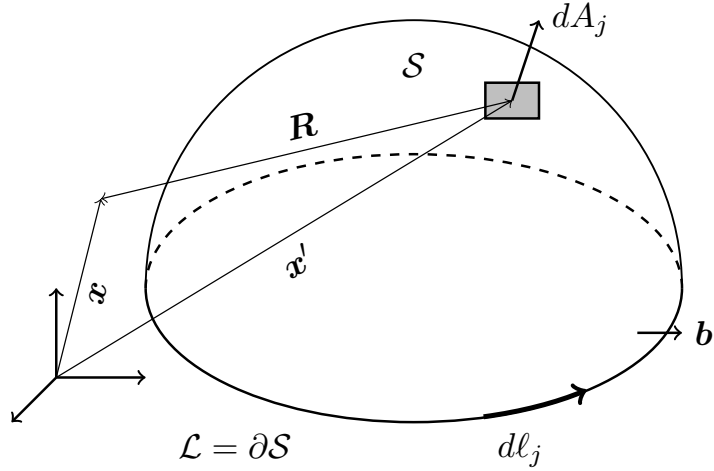} 
   \caption{The plastic distortion is concentrated on the surface $\mathcal{S}$, which is bounded by the dislocation line $\mathcal{L}=\partial\mathcal{S}$.}
   \label{fig:dislocation_loop}
\end{figure}
In this kinematic framework, dislocations are sources of plastic distortion (see Fig. \ref{fig:dislocation_loop}), and a particular form of the tensor $\bm\beta^P$ will be specified for both the classical and the gradient theory. Given a specific form of $\bm\beta^P$, another tensor of fundamental importance in the theory is the  dislocation density tensor $\bm\alpha$, which is obtained as the negative curl of $\beta^P_{ij}$, that is:
\begin{align}
\alpha_{ij}=-\epsilon_{jkm}\beta^P_{im,k}\, .
\end{align} 

\subsection{Classical anisotropic elasticity}
In classical anisotropic elasticity the strain energy density of the medium, $W$, is expressed as a quadratic form of the elastic distortions, that is 
\begin{equation}
W=\frac{1}{2}c_{ijkl}\varepsilon^E_{ij}\varepsilon^E_{kl}=\frac{1}{2}c_{ijkl}\beta^E_{ij}\beta^E_{kl},
\end{equation}
where $\varepsilon^E_{ij}$ is the symmetric part of $\beta^E_{ij}$ and $c_{ijkl}$ is the standard rank-4 tensor of elastic moduli. Given $W$ and the aforementioned kinematic assumption, the equilibrium equation, in the absence of body forces, becomes:
\begin{align}
\left(\frac{\partial W}{\partial\varepsilon^E_{ij}}\right)_{,j}=\sigma_{ij,j}=c_{ijkl}\left(u_{k,l}-\beta^{P0}_{kl}\right)_{,j}=L_{ik}u_k-c_{ijkl}\beta^{P0}_{kl,j}=0
\label{virtualWork_classical}
\end{align}
where $\sigma_{ij}={\partial W}/{\partial\varepsilon^E_{ij}}$ is the Cauchy stress tensor, and $L_{ik}=c_{ijkl}\partial_l\partial_j$ is the Navier differential operator. Eq.~(\ref{virtualWork_classical}) can be solved using the Green's function method.
For an infinite medium, the displacement and stress fields read, respectively \cite{Mura:1987wt}: 
 \begin{subequations}\label{eigen_solution0}
\begin{align}
u_{i}(\bm x)&= - c_{mnpq}G^0_{im,n}(\bm x)*\beta^{P0}_{pq}(\bm x)
\label{eigen_displ} \\
\sigma_{ij}(\bm x)&=c_{ijkl}\epsilon_{lqs}c_{mnpq}G^0_{km,n}(\bm x)*\alpha^0_{ps}(\bm x)\, .
\label{eigen_dist}
\end{align}
 \end{subequations}
 In eq.~(\ref{eigen_solution0}),  $G^0_{km}$ is the Green's function of the Navier operator and the symbol $*$ indicates convolution over three-dimensional space. The plastic eigendistortion corresponding to a Volterra dislocation extending over a surface $\mathcal{S}$   is:
 \begin{align}
\beta^{P0}_{kl}(\bm x)=-\int_\mathcal{S}\delta(\bm x-\bm x')b_k\, \text{d}A_l'
\label{dislocation_distortion_classical}
\end{align}
where $\delta$ is the Dirac delta function, and $b_i$ (Burgers vector) is the displacement jump across $\mathcal{S}$. 
Because $\bm b$ is constant,  the dislocation density tensor turns out to be concentrated on  the closed line $\mathcal{L}=\partial\mathcal{S}$ bounding the surface $\mathcal{S}$ (i.e. the dislocation line): 
\begin{align}
\alpha_{ij}^0(\bm x)=-\epsilon_{jkm}\beta^{P0}_{im,k}=\oint_{\mathcal{L}}\delta(\bm x-\bm x')b_i\, \text{d}L_j'
\label{dislocation_density_tensor_classical}
\end{align}

Finally, substituting eqs.~({\ref{dislocation_distortion_classical}) and (\ref{dislocation_density_tensor_classical}) into (\ref{eigen_solution0}), we obtain the classical expressions for the displacement and stress fields of a dislocation loop, respectively:
 \begin{subequations}\label{eigen_solution}
\begin{align}
u_{i}(\bm x)&=  \int_\mathcal{S}c_{mnpq}G^0_{im,n}(\bm x-\bm x')b_p\, \text{d}A_q'
\label{eigen_displ} \\
\sigma_{ij}(\bm x)&=\oint_{\mathcal{L}}c_{ijkl}\epsilon_{lqs}c_{mnpq}G^0_{km,n}(\bm x-\bm x')b_p\, \text{d}L_s'\, .
\label{eigen_dist}
\end{align}
\end{subequations}
Notice that both these expressions involve the derivatives of the anisotropic Green tensor. The classical Green tensor for anisotropic materials was first obtained by Lifshitz and Rozenzweig \cite{Lifshitz_1947} and Synge \cite{Synge:1957aa} using the Fourier transform method, and it reads:
\begin{equation}
G_{ij}^0(\bm{r})=\frac{1}{8\pi^2r} \int_{0}^{2\pi}\hat{L}_{ij}^{-1}(\bm n(\phi))\, \text{d}\phi\,
\label{ASGF}
\end{equation}
where $\hat{L}_{ij}(\bm k)=c_{ijkl}k_kk_l$ is the Navier operator in Fourier space, and $\bm n$ is a unit vector on the equatorial plane normal to $\bm r$.  The classical Green's tensor function and its gradient are singular at the origin, and thus the singularity is inherited in all of the fundamental equations of classical dislocation theory.

\subsection{Gradient anisotropic elasticity of Helmholtz type}
Gradient elasticity of Helmholtz type \cite{Lazar:2005fl,Lazar_2013,Lazar:2014dy} is a simplified version of Mindlin's gradient elasticity \cite{Mindlin:1964tg,Mindlin:1972vp,Mindlin:1968uu} with only one gradient parameter, where the strain energy density of the medium, $W$, is expressed as a quadratic form of the elastic distortion and its gradient: 
\begin{align}
W=\frac{1}{2}c_{ijkl}\beta^E_{ij}\beta^E_{kl}+\frac{\ell^2}{2}c_{ijkl}\beta^E_{ij,m}\beta^E_{kl,m}
\label{strain_energy_helmholtz}
\end{align}
For details regarding the reduction of Mindlin's general theory to this form, the reader is referred to the work of Lazar and Maugin \cite{Lazar:2005fl} and Lazar and Po \cite{LazarPo_PLA2015, PoLazar_ANSGF_2014}. In Eq.~(\ref{strain_energy_helmholtz}), the characteristic length, $\ell$, is the only gradient parameter to be determined by atomistic calculations. Owing to this particular form of the strain energy density, the equation of mechanical equilibrium, in the absence of body forces, reads:
\begin{align}
\left(\frac{\partial W}{\partial\varepsilon^E_{ij}}-\left(\frac{\partial W}{\partial\varepsilon^E_{ij,m}}\right)_{,m}\right)_{,j}=\left(\sigma_{ij}-\tau_{ijm,m}\right)_{,j}=c_{ijkl}\partial_j\left(1-\ell^2\partial_m\partial_m\right)\left(u_{k,l}-\beta^P_{kl}\right)=L_{ik}Lu_k-Lc_{ijkl}\beta^P_{kl,j}=0
\label{virtualWork_gradient}
\end{align}
where  $\tau_{ijm}=\partial W/\partial\varepsilon^E_{ij,m}=\ell^2c_{ijkl}\varepsilon^E_{kl,m}=\ell^2\sigma_{ij,m}$ is the double stress tensor, and $L=1-\ell^2\partial_m\partial_m$ is the Helmholtz differential operator. Eq.~(\ref{virtualWork_gradient}) is formally similar to~(\ref{virtualWork_classical}), with the difference that the composed Navier-Helmholtz operator $L_{ik}L$ replaces the Navier operator, and $L\beta^P_{kl}$ replaces $\beta^{P0}_{kl}$.  Letting $G_{im}$ be the Green's tensor of the composed Navier-Helmholtz operator, this similarity allows us to easily find the  solution of~(\ref{virtualWork_gradient}) as:
\begin{subequations}\label{eigen_solution_gradient}
\begin{align}
u_{i}(\bm x)&=- c_{mnpq}G_{im,n}*L\beta^{P}_{pq}=- c_{mnpq}G_{im,n}*\beta^{P0}_{pq}=\int_\mathcal{S}c_{mnpq}G_{im,n}(\bm x-\bm x')b_p\ dA_q'
\label{displ_gradient_Lazar_unisotrpoc}\\
\sigma_{ij}(\bm x)&=c_{ijkl}\epsilon_{lqs}c_{mnpq}G_{km,n}*L\alpha_{ps}=c_{ijkl}\epsilon_{lqs}c_{mnpq}G_{km,n}*\alpha_{ps}^0=\oint_{\mathcal{L}}c_{ijkl}\epsilon_{lqs}c_{mnpq}G_{km,n}(\bm x-\bm x')b_p\ dL_s'
\label{stress_gradient_Lazar_unisotrpoc}
\end{align}
\end{subequations}
Notice that in order to obtain the second equality in both~(\ref{displ_gradient_Lazar_unisotrpoc}) and~(\ref{stress_gradient_Lazar_unisotrpoc})  we have considered a plastic  eigendistortion spread according to
 \begin{align}
 \beta^P_{pq}=\beta^{P0}_{pq}*G
\label{plastic_distortion_gradient}
 \end{align} 
 so that $L\beta^P_{pq}=\beta^{P0}_{pq}$ and $L\alpha_{pq}=\alpha^{0}_{pq}$.

In order to implement Eq.~(\ref{stress_gradient_Lazar_unisotrpoc}) numerically, it is necessary to know the analytical expression of the Green's tensor function of the Helmholtz-Navier operator. This was recently derived by Lazar and Po \cite{PoLazar_ANSGF_2014} using the Fourier transform method and it consists of the convolution between the Green's tensors of the individual operators. It reads:
\begin{equation}\label{ANSGF}
G_{ij}(\textbf{x})=G^0_{ij}*G=\frac{1}{8\pi^2 \ell}\int_0^{2\pi}\int_0^1 L_{ij}^{-1}(\textbf k)\ \text e^{-Rq/\ell} \text dq\, \text d\phi,
\end{equation}
where the constant 1/$\ell$ term in front of the integral replaces the singular 1/$r$ term in equation (\ref{ASGF}). The gradient of the non-singular Green's tensor function of the Helmholtz-Navier operator reads \cite{PoLazar_ANSGF_2014}:
\begin{equation}\label{ANSGFD}
G_{ij,m}(\textbf{x})=-\frac{1}{8\pi^2 \ell^2}\int_0^{2\pi}\int_0^1 L_{ij}^{-1}(\textbf k)\  k_m\  \text e^{-Rq/\ell} \text dq\, \text d\phi,
\end{equation}
where $q=\cos \theta$ and $\text dq=-\sin \theta \text{d} \theta$ (see left side of figure \ref{dislCalcs}). In gradient anisotropic elasticity of Helmholtz type, both the Green's tensor function and its gradient are non-singular. However, compared to their singular counterparts, equations (\ref{ANSGF}) and (\ref{ANSGFD}) require an additional integral over the azimuthal angle in Fourier space.  

\subsection{Atomistic calculations of stress}
In order to justify the comparison between continuum and atomic stresses, we shall now discuss the definition of the latter. The equivalent atomistic description of the continuum Cauchy stress has for some time been the subject of debate \cite{Zhou_2003, Zimmerman_2004, Subra_2008}. Being a concept derived and understood at the continuum level, complications arise in the calculation of atomic stresses, primarily when finite temperatures are involved. Such complications stem from uncertainties in the definition of the volumes enclosing the material point of interest and how to adequately average atomic forces and velocities both spatially and temporally. 
Using a statistical mechanics approach, where the stress is described by the local momentum flux through an enclosed surface, Lutsko \cite{Lutsko_1988} developed a microscopic stress tensor suitable for molecular simulations. Cormier et al. \cite{Cormier_2001}, expanded this approach and showed its validity near highly strained regions in the material by comparing with the solutions of classical anisotropic elasticity theory.  While several additional schemes have also been developed to treat various calculations \cite{Tsai_1979, CheungYip_1991, Hardy_1982}, the virial stress, based on the virial theorem of gases by Clausius \cite{Clausius_1870}, remains the most common method of calculation. In this representation, the general per-atom stress tensor for an atom $\alpha$, can be expressed as
\begin{equation}\label{virial}
\sigma_{ij}^{(\alpha)}=-\frac{1}{\Omega} \left [ m^{(\alpha)} v_i^{(\alpha)} v_j^{(\alpha)} +\frac{1}{2}\sum_{\beta=1}^{N}( r_i^{(\alpha)}- r_i^{(\beta)}) F_j^{(\alpha)(\beta)} \right ]
\end{equation}
where $m^{(\alpha)}$, $\bf v^{(\alpha)}$, and $\bf r^{(\alpha)}$ are the mass, velocity, and position of atom $\alpha$, and $\bf F^{(\alpha)(\beta)}$ is the force vector on atom $\alpha$ due to atom $\beta$. For bcc metals the atomic volume may be taken as $\Omega = a^3/2$, and the summation is taken over $N$ nearest neighbor atoms within a prescribed cutoff distance. For the static calculations ($T$= 0 K) performed in this work, the first term in (\ref{virial}) vanishes and thus no averaging schemes are required, as the equivalent Cauchy stress is directly obtained \cite{Subra_2008}.The force vector $\bf F^{(\alpha)(\beta)}$ in (\ref{virial}) is computed directly from the total energy, $U^{tot}$, as
\begin{equation}\label{atForce}
\textbf F^{(\alpha)(\beta)} = -\frac{\partial U^{tot}}{\partial r^{(\alpha)(\beta)}}\frac{\textbf{r}^{(\alpha)(\beta)}}{\| \textbf{r}^{(\alpha)(\beta)}\|}
\end{equation}
where $\textbf{r}^{(\alpha)(\beta)}=\textbf{r}^{(\alpha)}-\textbf{r}^{(\beta)}$. We note that equation (\ref{virial}) is not momentum conserving and may produce erroneous results (e.g. non-zero normal stress components at free surfaces) when used in certain cases \cite{CheungYip_1991, Zimmerman_2004}.

\section{Calculations\label{sec_Calculations}}
The anisotropic non-singular (A-NS) theory is enabled through the extraction of the characteristic length parameter which we obtain through a fitting procedure with the results of atomistic calculations. In this section we describe how these calculations were performed and the procedure by which the length parameter was subsequently obtained.
\subsection{Atomistic Calculations}
Molecular statics (MS) relaxations of 1/2$\langle$111$\rangle$ screw and 1/2$\langle$111$\rangle\{$110$\}$ edge dislocations in iron were performed. The atomic interactions were treated with a recently developed magnetic bond-order potential (BOP) for iron by Mrovec et al. \cite{Mrovec_2011}. Based on the tight-binding approximation, bond-order potentials \cite{Pettifor_1989} offer a link between electronic structure and traditional molecular treatments. This potential also incorporates the Stoner model of itinerant magnetism to capture the effects of magnetic interactions between iron atoms. More commonly used interatomic potentials for iron such as that of Ackland et al. \cite{Ackland_2004}, struggle to reproduce important directional bonding characteristics and are not capable of addressing the issue of magnetism, which in iron has important consequences. Furthermore, such potentials have been specifically known to poorly reproduce such properties as the Peierls barrier for the motion of $1/2 \langle111\rangle$ screw dislocations compared to density functional theory \cite{Ventelon_2007} and experimental results. In contrast, the present BOP has been shown to very accurately reproduce bonding characteristics and dislocation properties in iron.

A schematic of the simulation cell used in the atomistic calculations is shown to the right in figure \ref{dislCalcs} with orientation and geometric parameters listed in table \ref{paramTable}. 
\begin{figure}[t!]
   \centering
   \includegraphics[width=.59\textwidth]{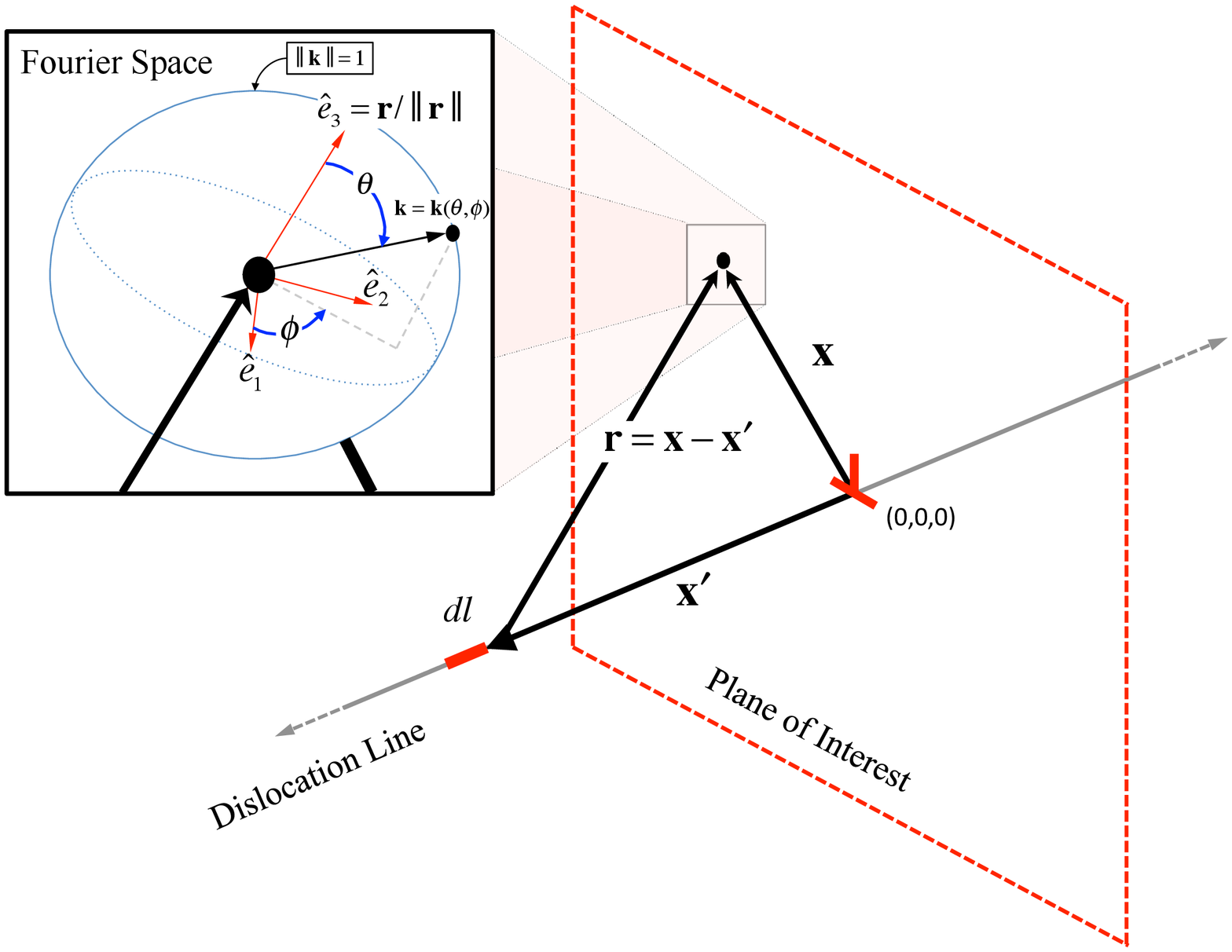} 
    \includegraphics[width=.4\textwidth]{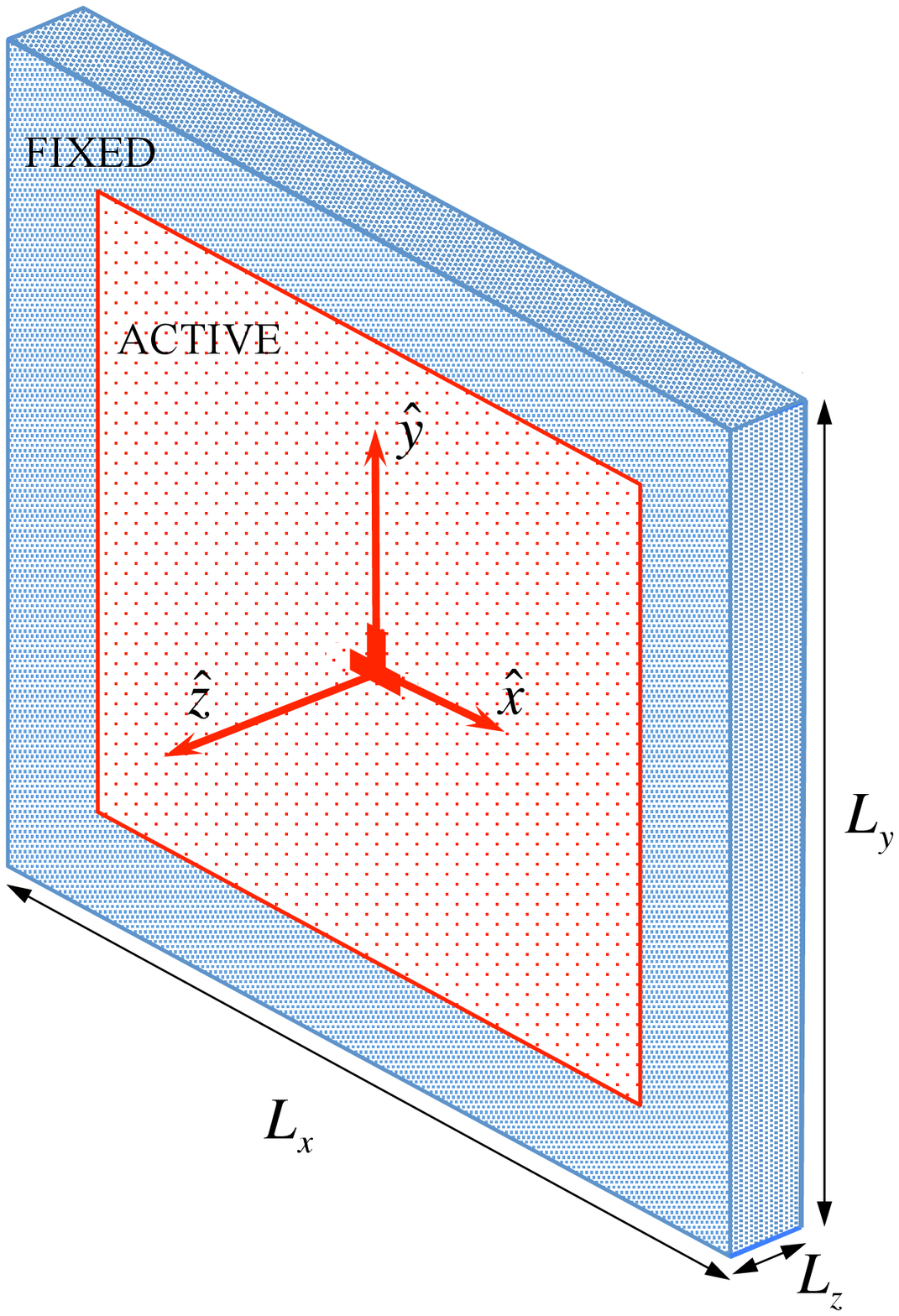} 
   \caption{Schematic of the geometries used in the continuum stress calculations (left) and the atomistic simulations (right).}
   \label{dislCalcs}
\end{figure}
The dislocations, both with line directions along the (periodic) $z$ direction, were introduced in the center of the box by displacing all atoms according to the anisotropic singular (A-S) elastic solutions (equation (\ref{ASGF})). Atoms in the fixed region were then held rigid, while atoms in the 'active' region were allowed to relax according to the Fast Inertial Relaxation Engine (FIRE) minimization scheme \cite{Bitzek_2006}. The relaxations were considered ``complete'' when the largest force on any atom is less than 0.01 eV/$\AA$. Stresses in this relaxed configuration were then computed according to (\ref{virial}).
\begin{table}[t]
\begin{center}
\begin{tabular}{cccccccc}
\hline
 &$\hat x$&$\hat y$&$\hat z$&$L_x/a$&$L_y/a$&$L_z/a$&Active(Fixed) Atoms\\
 \hline \hline \vspace{-.4cm}\\
1/2$\langle$111$\rangle$ screw dislocation               &$[\bar 1 \bar 12]$  &$[1 \bar 1 0]$    &$[111]$             &92      &92   & $\frac{\sqrt{3}}{2}$               & 11550(7344)                \\
1/2$\langle$111$\rangle\{$110$\}$ edge  dislocation &$[111]$                  &$[1 \bar 1 0]$    &$[11 \bar 2]$     &110    &108 & $\sqrt{6}$ & 55056(3498)                \\
\hline
\end{tabular}
\caption{Orientation and geometric parameters used in the atomistic calculations (lattice parameter $a = 2.85 \AA$).}\label{paramTable}
\end{center}
\end{table}

\subsection{Extraction of the Characteristic Length}
The characteristic length $\ell$ in the gradient elasticity formulation equations (\ref{strain_energy_helmholtz}), (\ref{ANSGF}), and (\ref{ANSGFD}) may be obtained once the BOP relaxations have been performed. To accomplish this, a fitting procedure based on the minimization of an objective function is implemented. The objective function is the sum of the spectral norms of the residual matrices, denoted as SSNR$^{(\ell)}$, over a chosen group of $N$ ($\sim$16-24) atomic positions ($\textbf r^{(\alpha)}$) surrounding the dislocation core. The residual matrix represents the difference of the calculated stress tensors from the non-singular continuum theory,~$\pmb{\sigma}^{(\ell)}$, and that from the atomistics,~$\pmb{\sigma}^{(\text{BOP})}$. Thus, the objective function can be expressed as
\begin{equation}\label{SSNR}
\text{SSNR}^{(\ell)}=\frac{1}{\mu}\sum_{\alpha=1}^{N} \left \|  \pmb{\sigma}^{(\ell)}(\textbf r^{(\alpha)})-\pmb \sigma^{\text{BOP}}(\textbf r^{(\alpha)}) \right \|,
\end{equation}
and is normalized by the shear modulus, $\mu$, where $\mu=c_{44}$. The optimal characteristic length parameter can then be found as the value of $\ell$ that minimizes (\ref{SSNR}). 

To better understand the role of each stress component in determining $\ell$, a matrix of the sum of the norm of the residual components, denoted as SNR$_{ij}^{(\ell)}$, was calculated as 
\begin{equation}\label{SNR}
\text{SNR}_{ij}^{(\ell)}=\frac{1}{\mu}\sum_{\alpha=1}^{N} \left \|  \sigma_{ij}^{(\ell)}(\textbf r^{(\alpha)})-\sigma_{ij}^{\text{BOP}}(\textbf r^{(\alpha)})  \right \|.
\end{equation}
A plot of the components of $\text{SNR}_{ij}$ versus $\ell$ will reveal the sensitivities and dependencies of $\ell$ in reproducing the atomistic stresses, as will be seen in the section \ref{sec_Results}.

\subsection{Continuum Stress Calculations}
A schematic of the geometry used for the continuum calculations is given in the left side of figure \ref{dislCalcs}. For comparison, calculations are also performed using the singular anisotropic (A-S) and isotropic (I-S) solutions from the classical approach. The calculations are performed at the reference atomic positions super-imposed onto a plane normal to the dislocation line. To calculate the stresses of an infinitely long straight dislocation, we use the loop equations for stress ((\ref{eigen_dist}) and (\ref{stress_gradient_Lazar_unisotrpoc})), letting the loop radius go to infinity. Thus we must choose an appropriately long dislocation line such that the contributions at and beyond the endpoints on the line are negligible. A dislocation line length of 30$b$ was found to be sufficient, where $b$ is the magnitude of the Burgers vector. The plane where the calculations were performed was placed intersecting the midpoint of the dislocation line. The line was then populated with 50 evenly spaced points for the numerical integration of the stress equation. For each integration step along the line, integrations of the Green's tensor function are required. These Fourier-space integrations are performed using Gaussian quadrature integration on the unit circle lying on the plane perpendicular to $\bf{\hat e_3}$ (A-S) and over the unit hemisphere (A-NS). For the former, 8 quadrature points were found to be sufficient, while for the latter, the most critical calculation required 30 and 200 points for the integrations over $\phi$ and $q$, respectively. The elastic constants obtained from the iron potential ($c_{11}=243$ GPa, $c_{12}=145$ GPa, and $c_{44}=\mu=119$ GPa) were used in all elasticity calculations. We note that for the isotropic-singular (I-S) solutions, the same methodology can be followed as the A-S with the substitution of $c_{11}\rightarrow 2c_{44}+c_{12}$.

\section{Results \& Discussion\label{sec_Results}}
\subsection{Screw Dislocation}\label{ResScrew}
Figure \ref{LFits} plots the results of the optimization of $\ell$ for both screw (left) and edge (right) dislocations. Components of the sum of the normed residual ($\textbf{SNR}^{(\ell)}$) matrix versus characteristic length parameter are plotted with inset plots showing the sum of the spectral norm of the residual matrices, $\text{SSNR}^{(\ell)}$, whose minimum determines the overall best choice of $\ell$. For the screw dislocation, the minimums of the $yz$ and $xz$ components of the $\textbf{SNR}^{(\ell)}$ matrix give values of $\ell$ of 0.232 $a$ and 0.316 $a$, respectively. In the inset plot of $\text{SSNR}^{(\ell)}$ the minimum is found at a value of $\ell=$0.288 $a$, which we take as the best fit value for $\ell$.
%
\begin{figure}[h!]
   \centering
   \includegraphics[width=1\textwidth]{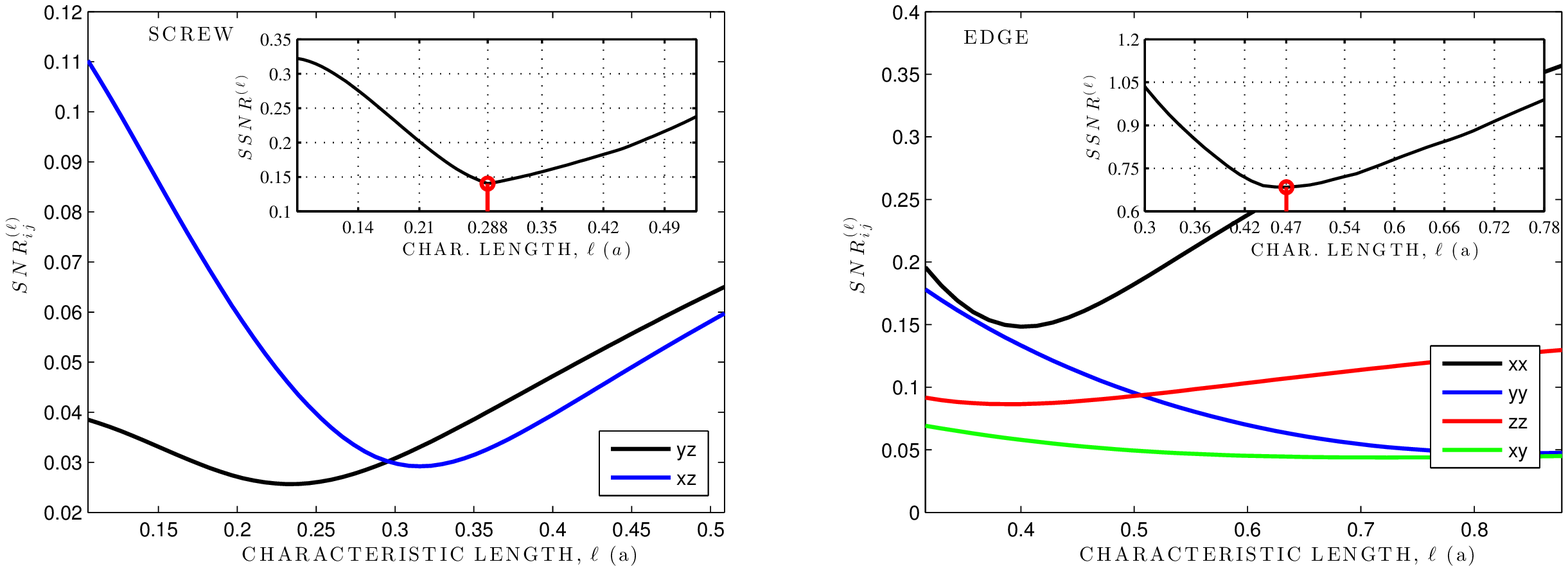} 
   \caption{Plots of the components of the sum of the normed residuals ($\textbf{SNR}^{(\ell)}$) versus characteristic length parameter for screw (left) and edge (right) dislocations. The minimum of each component gives the best value of $\ell$ considering only that component. The inset plots show the sum of the spectral norm of the residual matrices, $\textbf{SSNR}^{(\ell)}$, whose minimum (encircled) determines the overall best choice of $\ell$, considering the contributions of all components.}
   \label{LFits}
\end{figure}
%
In figure \ref{screwS}, the results of the BOP relaxations and continuum calculations are plotted for two rows of atoms below and two rows above the dislocation core in the [1$\bar{1}$0] direction. In each plot, the abscissa represents the [$\bar{1}\bar{1}$2] direction in units of lattice parameters. It is immediately clear that the isotropic singular (I-S) solution gives a drastic over-estimation of the atomistic calculations, most notably in the $xz$ component, where errors as high as 3-4 times are observed. The A-S solution, however, yields stress values remarkably close to the BOP calculations, within several percent, with the exception of the $xz-$component of the atoms belonging to the core. Results of the A-NS calculations are plotted as green points (where $\ell$=0.288 $a$) on a vertical line whose endpoints have horizontal bars representing the lower (black) and upper (red) values of $\ell =$ 0.232 $a$ and 0.316 $a$, inferred from the range of minima of the components of $\textbf{SNR}_{ij}^{(\ell)}$ in figure \ref{LFits}. The purpose of this line is to illustrate the sensitivity of the stress calculations to variations in the chosen $\ell$, which in this case we find to be negligible. For the screw dislocation, the A-S and A-NS solutions both provide excellent agreement with the BOP stresses for all atomic locations except for the core's 3 nearest neighbor atoms, where A-NS gives values of 0-22\% larger than the atomistic calculations and is thereby much closer to atomistics than the singular fields which deviate by 65-103\%.

\begin{figure}[h!]
   \centering
   \includegraphics[width=.9\textwidth]{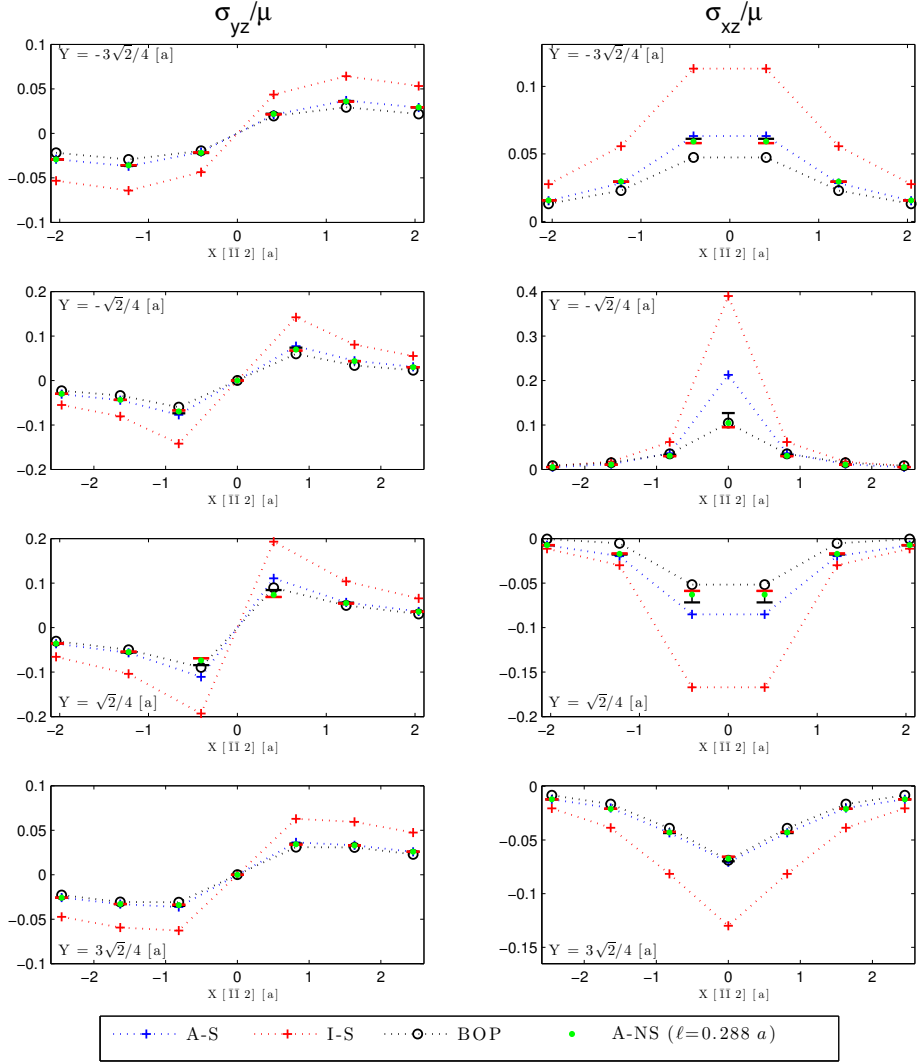} 
   \caption{Normalized stress calculations for a screw dislocation at atomic locations two rows above ($Y$=$\sqrt{2}/4$$a$, $3\sqrt{2}/4$$a$) and two rows below the dislocation core (Y=-$\sqrt{2}/4$$a$, -$3\sqrt{2}/4$$a$) in the [1$\bar 1$0] direction. In each plot, the abscissa, $X$, represents the [$\bar1\bar1$2] direction, and the dislocation core is centered at $X$=0, $Y$=0. Plotted are the singular anisotropic and isotropic elastic solutions (A-S and I-S), the results from the atomistic simulations (BOP), and the anisotropic non-singular elastic solutions (A-NS) using the best fit value of $\ell = 0.288 a$. Horizontal black and red line segments represent A-NS stress calculations performed using the bounding values of $\ell$ taken from figure \ref{LFits}, as discussed in section \ref{ResScrew}, and are connected with a vertical line to aid in visualizing. Black and red segments represent stress calculations using $\ell$ = 0.232 $a$ and 0.316 $a$, respectively.}
   \label{screwS}
\end{figure}
\subsection{Edge Dislocation}\label{ResEdge}
For the edge dislocation, the deviations between the components of the $\textbf{SNR}^{(\ell)}$ matrix in figure \ref{LFits} are found to be larger, ranging from a minimum of $\ell = $0.386 $a$ ($zz$ component) to a maximum $\ell = $0.860 $a$ ($xy$ component).  While this represents quite a large range of candidate values for $\ell$, we find that due to its large sensitivity to changes in $\ell$ and its highly localized minimum near $\ell = $0.4 $a$, the $xx$ component of $\textbf{SNR}^{(\ell)}$ has the most influence on the minimum of $\textbf{SSNR}^{(\ell)}$, giving it a well-defined minimum at the best fit value of 0.470 $a$.
The results of the BOP relaxations and continuum calculations are plotted in figure \ref{edgeS}, for the two atomic rows below, and two atomic rows above the glide plane in the [1$\bar{1}$0] direction.                         
\begin{figure}[h!]
   \centering
   \includegraphics[width=1\textwidth]{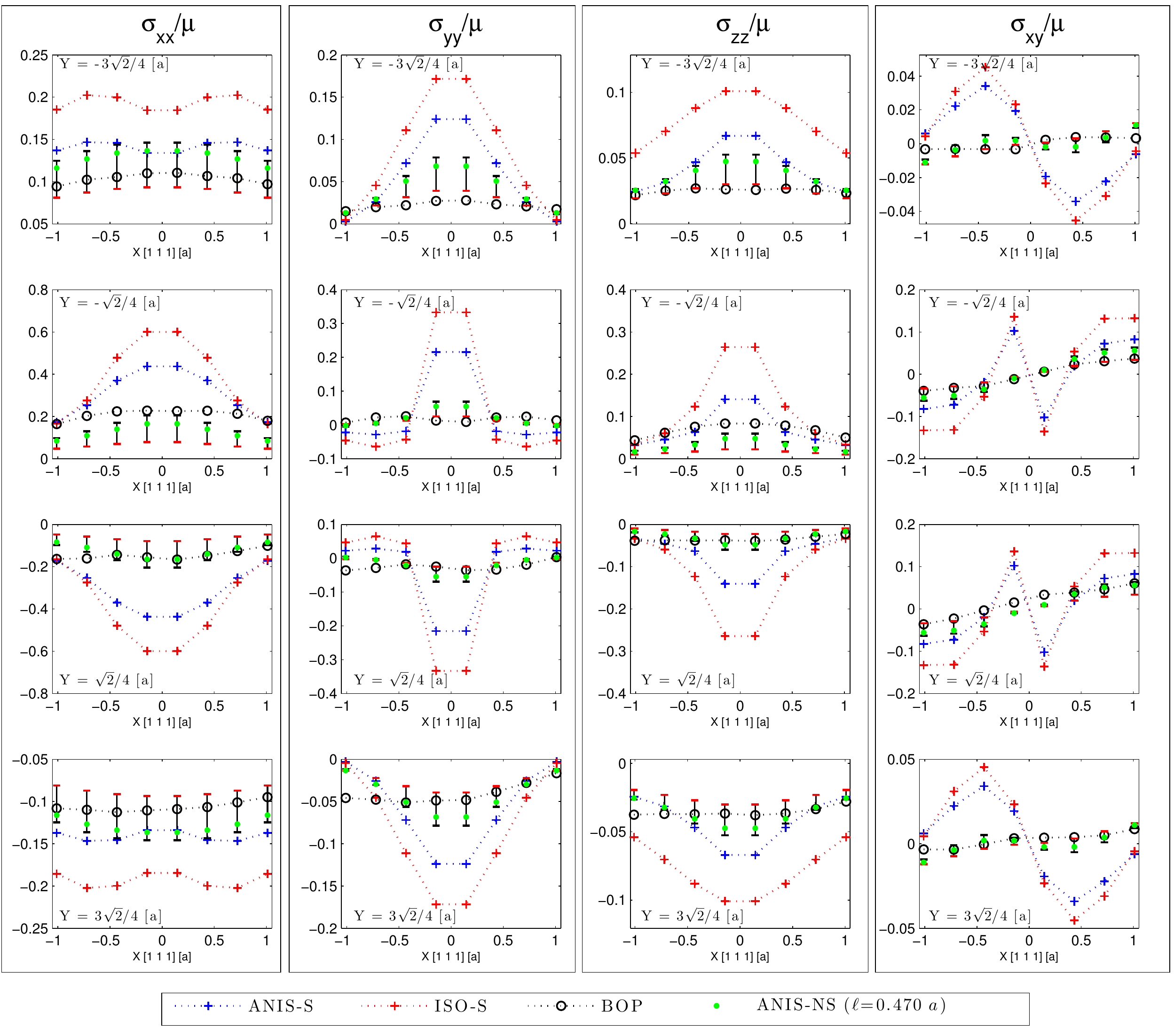} 
   \caption{Normalized stress calculations for an edge dislocation at atomic locations two rows above ($Y$=$\sqrt{2}/4$~$a$, $3\sqrt{2}/4$~$a$) and two rows below the glide plane (Y=-$\sqrt{2}/4$~$a$, -$3\sqrt{2}/4$~$a$). In each plot, the abscissa, $X$, represents the Burgers direction, and the dislocation core is centered at $X$=0, $Y$=0. Plotted are the singular anisotropic and isotropic elastic solutions (A-S and I-S), the results from the atomistic simulations (BOP), and the anisotropic non-singular elastic solutions (A-NS) using the the best fit value of $\ell = 0.470 a$. Connected horizontal black and red line segments represent A-NS stress calculations evaluated at $\ell$ = 0.386 $a$ and 0.860 $a$, respectively.}
   \label{edgeS}
\end{figure}
In each plot, the abscissa represents the [111] (Burgers) direction in units of lattice parameters. For the edge dislocation, the results show the BOP relaxations differ significantly from both of the I-S and A-S solutions. As expected, the I-S solutions again give a drastic over-estimation of each component of stress, in some instances nearly an order of magnitude too large. The A-S solutions offer some improvement, keeping the same trends of the I-S solutions, but decreasing their magnitude by typically between 25-50$\%$. The trends predicted by the singular theories, however, are not observed in these atomic rows nearest to the glide plane. This is where we find the greatest benefit in implementing the non-singular solution. In figure \ref{edgeS}, the A-NS solutions are presented as green points (where $\ell =$ 0.470  $a$) on a vertical line that goes from $\ell$ = 0.386 $a$ (black bar) to 0.860 $a$ (red bar), taken from figure \ref{LFits}. At the two nearest rows to the glide plane, we see very good agreement with the BOP calculations, for every stress component. The very small-magnitude sinusoidal behavior of the $xy$ component is also very nicely captured, where the singular solutions have failed. One notable characteristic for the edge dislocation is the much larger variations in the calculated stresses over the range of candidate $\ell$'s used. This is most prominent in the hydrostatic stress components in the $Y = \pm 3\sqrt{2}/4~a$ atomic rows. The A-NS solutions predict a weaker decay in stress than observed in the atomistics, suggesting improvement with a larger value of $\ell$. However, the differences compared to the BOP calculations are always relatively small compared to the singular cases which can be off by a factor of greater than an order of magnitude. A unique aspect worth noting was the ability of the A-NS calculations at multiple atomic locations to give the correct sense of the atomic stresses, where the singular solutions failed. This is seen in the atomic rows on the glide plane in $\sigma_{yy}$, and for some positions in $\sigma_{xy}$. 

\subsection{Practical implications}
The results presented thus far have illustrated the drastic increase in accuracy in implementing the anisotropic non-singular solutions near the dislocation core. However, since this additional accuracy comes at a computational cost due to the additional nested integral in the non-singular Green's function, in practice it is vital to know the spatial extent to which the A-NS solutions are necessary and where the singular solutions are sufficient. To do this, we have plotted in figure \ref{PEtot2} a percent difference field of the A-S stress tensor with the A-NS stress tensor for a screw dislocation (left) and and edge dislocation (right), according to the expression 
\begin{equation}\label{PDIFF}
\% \text{DIFF}(\textbf r)^{\text{A-S}} = 100\%\cdot \frac{\left \|  \pmb{\sigma}^{\text{A-S}}(\textbf r)-\pmb \sigma^{\text{A-NS}}(\textbf r) \right \|}{\left \| \pmb \sigma^{\text{A-NS}}(\textbf r) \right \|}.
\end{equation}
In the figure, an overlaid grid of points represent a reference atomic lattice for illustrative purposes. Additionally, all locations with a percent difference greater than 100 and 250 are lumped into the same color for the edge and screw dislocation, respectively. 
\begin{figure}[h!]
   \centering
   \includegraphics[width=.49\textwidth]{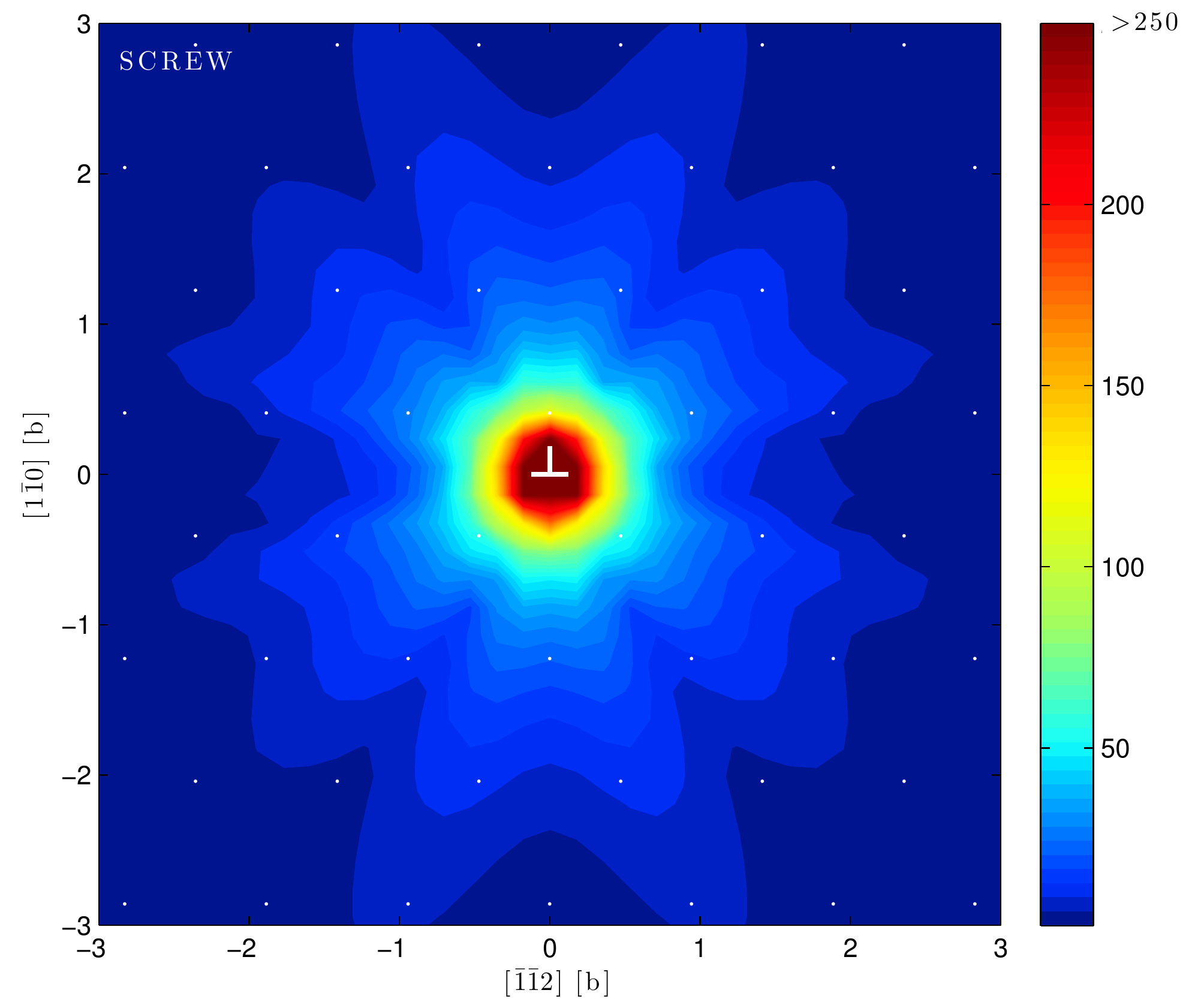}
     \includegraphics[width=.49\textwidth]{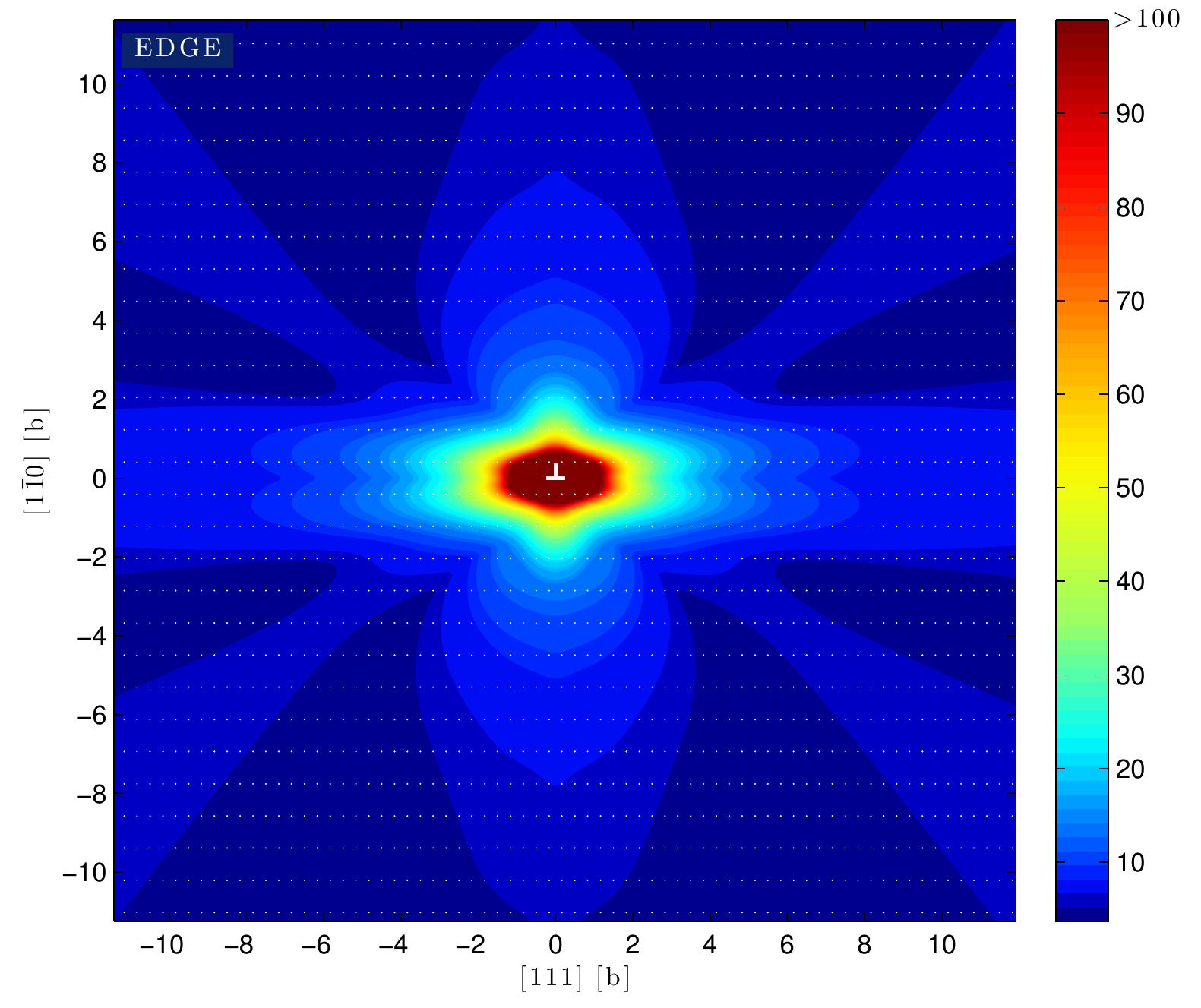}   
   \caption{Percent difference field of the A-S stress tensor with the A-NS stress tensor (calculated from (\ref{PDIFF})) for a 1/2$\langle$111$\rangle$ screw dislocation (left) and a 1/2$\langle$111$\rangle\{$110$\}$ edge dislocation (right). Corresponding reference atomic lattices are super-imposed for convenience.}
   \label{PEtot2}
\end{figure}
For the screw dislocation, the A-S and A-NS solutions converge to less than 20$\%$ and 3$\%$ difference at radial distances of approximately $0.9~b$ to $3~b$ from the core, respectively.  For the edge dislocation, the differences extend to a much longer range in the material. Along the atomic rows adjacent to the glide plane, the A-S differs from the A-NS solution by $\geq 100\%, 50\%, 20\%, 10\%, $ at $|x| \leq 1.0~b, 1.9~b, 3.3~b,$ and $6~b$, respectively. Along a line perpendicular to the glide plane and intersecting the core, the A-S differs from the A-NS solution by $\geq 100\%, 50\%, 20\%, 10\%, $ at $|y| \leq 0.65~b, 1.0~b, 2.2~b,$ and $4.4~b$, respectively. 

Similar calculations of the percent difference field of the I-S solutions with the A-NS show a convergence to a (roughly) constant difference of $\approx 85\%$ occurring just beyond the core atoms for the screw dislocation. For the edge dislocation, along the atomic rows adjacent to the glide plane, the I-S differs from the A-NS solution by $\geq 100\%$ and $50\%$ at $|x| \leq 1.7~b, 4.0~b$, respectively. At larger distances, the differences oscillate between a minimum of $\approx 20~\%$ to a maximum of $\approx 50~\%$ for $\sqrt{x^2 + y^2} > 4 b$.

\subsection{General discussion}

Implementation of the non-singular approach requires a priori knowledge of the characteristic length for the material. While in this work we have focused solely on near-core dislocation stress fields from atomistic calculations as means to extract this material parameter, there are a variety of general approaches that one may follow to obtain it. Using density functional theory (DFT), Shodja et al. \cite{Shodja_2013} were able to calculate strain-gradient characteristic lengths for several nearly isotropic materials ($A~\approx$ 1), by relating the calculated Hessian matrix of the system to the double stress tensor. Other approaches include the development of analytical expressions derived from interatomic potentials \cite{Shodja_2010}. In our approach, we found that fitting to the fields of both screw and edge dislocations did not result in a single characteristic length. On the contrary, we found $\ell^{edge}$ to be roughly 63\% larger than $\ell^{screw}$. However, noting the extremely minute variations of the stresses with changes in $\ell$ seen in figure \ref{screwS}, we can easily justify the use of the length parameter obtained from the edge dislocation ($\ell$=0.470 $a$) to describe both dislocations, and thus such a discrepancy poses no conflict in the case of iron. For other materials, this aspect remains an outstanding issue.

Current methods of point defect and dislocation simulations stand to benefit greatly from the increased accuracy that the non-singular implementation provides. In DD simulations, ad hoc assumptions of the short range interaction of dislocations may now be replaced with non-singular fields based on and verified by atomistic calculations. This includes a possible re-definition of the dislocation self force that may account for near-core contributions and material anisotropy, where the line-tension approximation is insufficient. Alternatively, simplified descriptions of near core behavior in DD simulations may be calibrated to match the non-singular solution for the fields near the dislocation core. Mechanisms of dislocation junction formation as well as pinning and de-pinning and interactions with precipitates may also be better understood when such contributions are accounted for. In the same way as demonstrated in this work, the method could be implemented for the other transition metals and thereby help to analyze and possibly understand the differences of the bcc transition metals with respect to anomalous slip \cite{Christian_1983,Marichal_2014}.

The diffusion of point defects in dislocation stress fields is known to be anisotropic due to drift forces acting on the point defect resulting from the interaction of the dislocation stress field and the point defect's formation volume. In simulations of this type, the large overprediction of the dislocation stresses by the singular theories leads to large overpredictions of the interaction energy, and thus, important material parameters such as dislocation capture rates for interstitials and vacancies are inherently unreliable. Among other issues, this has important consequences for simulations of creep and swelling behavior in irradiated metals. In the case of the dislocation bias factor \cite{Wolfer_2007}, a weaker interaction field may explain the order of magnitude overestimation of bias factors calculated with elasticity theory compared to those expected from empirical swelling data.

\section{Summary \& Conclusions\label{sec_Conclusions}}

An anisotropic non-singular continuum theory of dislocation fields was derived and applied to accurately represent the near-core stresses of dislocations in iron. The theory was enabled through the use of the recently developed non-singular Green's tensor function of gradient anisotropic elasticity of Helmholtz-type \cite{PoLazar_ANSGF_2014}, which requires only a single characteristic length parameter in addition to the material's elastic constants. Using a magnetic bond-order potential to model iron interactions, molecular statics calculations were performed, and a fitting procedure was developed to extract the optimum length parameter required by the non-singular theory. Results show the method can accurately replicate the magnitude and decay of the near-core dislocation stresses even for atoms belonging to the core itself. Comparisons with the singular isotropic and anisotropic theories show the non-singular anisotropic theory leads to a substantially more accurate representation of the stresses of both screw and edge dislocations near the core. 

\noindent Conclusions of our study can be summarized as follows:

\begin{itemize}
\item We have shown that an anisotropic non-singular expression for the stress as a line integral may be obtained by direct substitution of the anisotropic non-singular Green's tensor function gradient of the Helmholtz-Navier type \cite{PoLazar_ANSGF_2014} into the classical line integral expression.
\item A fitting procedure was developed and shown to be an adequate methodology for the extraction of the material length parameter, $\ell$, required by the non-singular theory. In bcc iron, this parameter is found to be $\ell=0.288~a$ and $0.470~a$ when fitted to the atomic stress fields of screw and edge dislocations independently, respectively. Due to the marginal sensitivity of calculated stresses on $\ell$ for the screw dislocation, it is sufficient to take the value of $0.470~a$, extracted from the edge dislocation calculations, as a single, best value for $\ell$.
\item For screw dislocations, the non-singular stresses match extremely well with the BOP calculations, even including the core atomic locations, where the singular solutions over-predict stresses by factors of greater than two.
\item For edge dislocations, the non-singular solutions offer a drastic improvement in matching the BOP calculations over the singular solutions, where stresses are found to be as high as an order of magnitude too large in some cases. At multiple locations, the non-singular solutions are also found to match the correct sign of the atomic stresses, where the singular solutions give incorrect signs.
\item In practice, for an arbitrary dislocation loop or segment, it is advised that if less than 10$\%$(50$\%$) error with the non-singular solution is desired, the anisotropic-singular solutions should be implemented no farther than a cut-off radius of $\approx$6$b$(1.9$b$) from the core. For similar cut-off distances, the isotropic singular solutions oscillate between $\approx$20-85$\%$ error, depending on the angle relative to the core and the character (screw, edge, mixed) of the current segment being integrated. 
\end{itemize}

The results of this study highlight an advance in the capabilities of continuum elasticity approaches to resolve material stresses at sub-nanometer distances from dislocations, typically reproducible only by atomistic simulations. The general procedure described may in principle be applied to model any cubic anisotropic material where high accuracy calculations of near-core dislocation stresses are required. 

\section*{Acknowledgements}
D.S. would like to acknowledge support from the Alexander von Humboldt Foundation for a postdoctoral fellowship and research stay at Fraunhofer IWM in Freiburg. M.L. and P.G. gratefully acknowledge grants from the Deutsche Forschungsgemeinschaft (Grant Nos. La1974/2-2, La1974/3-1, Gu367/36).

%

\end{document}